\documentclass[]{spie}  

\usepackage{amsmath,amsfonts,amssymb}
\usepackage{graphicx}
\usepackage[colorlinks=true, allcolors=blue]{hyperref}
\usepackage{multirow}  
\usepackage{eurosym}
\usepackage[colorlinks=true, allcolors=blue]{hyperref}

\title{FATE - an operational automatic system for optical turbulence forecasting at the Very Large Telescope}

\author[a]{Elena Masciadri}
\author[a]{Alessio Turchi}
\author[a]{Luca Fini}
\author[b]{Alberto Ortolani}
\author[b]{Valerio Capecchi}
\author[b]{Francesco Pasi}
\author[c]{Angel Otarola}
\author[c]{Steffen Mieske}

\affil[a]{INAF-Osservatorio Astrofisico di Arcetri, L.go Enrico Fermi 5, Firenze, Italy}
\affil[b]{LaMMA, Via Madonna del Piano 10, Sesto Fiorentino, Firenze, Italy}
\affil[c]{ESO, Alonso de C\'ordova 3107, 763000, Santiago, Chile}

\authorinfo{Send correspondence to Elena Masciadri - e-mail: elena.masciadri@inaf.it}

\begin{document} 
\maketitle

\begin{abstract}
In this contribution we report the on-going progresses of the project FATE, an operational automatic forecast system conceived to deliver forecasts of a set of astroclimatic and atmospheric parameters having the aim to support the science operations (i.e. the Service Mode) at the Very Large Telescope. The project has been selected at conclusion of an international open call for tender opened by ESO and it fits with precise technical specifications. In this contribution we will present the ultimate goals of this service once it will be integrated in the VLT operations, the forecasts performances at present time and the state of the art of the project.  FATE is supposed to draw the roadmap towards the optical turbulence forecast for the ELT.
\end{abstract}


\section{INTRODUCTION}
\label{sec:intro}  

FATE is an acronym that states for Forecast system for Atmosphere and Turbulence for ESO. It is the winner project of an international Call for Tender (CfT) opened by ESO on 2020. The goal of the CfT was to supply automatic forecasts of optical turbulence and atmospheric parameters in operational mode for the VLT. The ESO request was to provide forecasts in operational mode for two years with the possible extension of up to further three years. The main goals/objectives of the ESO CfT were: \\
\begin{itemize}
\item to decreases the amount of out-of-constraints observations due to unforeseen changes in meteorological or optical turbulence conditions

\item to assist the support astronomers in making decision at night

\item to enable more aggressive short-term scheduling of the observations with well understood risks

\item to best prepare the mode operations of the ELT at Cerro Armazones by gaining experience on the use of the forecast to maximize the science return of the VLT at Cerro Paranal.

\end{itemize}

In other words a two fold objective: to retrieve forecast of the main parameters (atmospheric and optical turbulence) to support the Service Mode at the VLT\cite{silva2002,anderson2024} but also to prepare the roadmap towards the ELT. Since more than a decade ESO applies the Short-Term Scheduling (STS) whose aims is that to optimise the calculation of how many highly ranked observations are completed for unit time. Forecasts at short time scales are therefore useful to such an optimisation.

The roles and the actors inside FATE are as in the following: (1) ESO is the customer, (2) INAF is the contractor with the PI-ship, (3) the FATE team is composed by two institutes: INAF and LaMMA, both located in Florence, Italy. \\ \\

INAF have a robust and long-time expertise in the optical turbulence forecast field and atmospheric parameters relevant for the ground-based astronomy. The most relevant progresses in the context of the optical turbulence forecast have been reached in the proposition of parameterisation solutions, charaterisation of sites, proposition of model calibration methods hybrid forecast methods\cite{masciadri1999,masciadri2001,masciadri2006,lascaux2011,hagelin2011,masciadri2013,masciadri2017,masciadri2020,turchi2019,turchi2020,masciadri2023}. 
LaMMA has a long-time experience in managing operational meteo services and it is the meteo structure supporting the Civil Protection at an italian regional level that is in Tuscany.\\

INAF has in FATE the following responsibilities: (1) scientific responsibility of the project,  (2) development of the optical turbulence model Astro-Meso-Nh\cite{masciadri1999} and development of the automatic forecast system\cite{turchi2024}, (3) development of algorithms having the aims to achieve the technical specifications and (4) R\&D to further improve forecast performances. \\
LaMMA has the responsibility to manage FATE forecasts in an operational mode on a daily base. \\ 

We highlight that INAF had carried out in the past a feasibility study (called MOSE - Phase A (2011-2013) and Phase B (2014-2015)) for ESO aiming to test the possibility to provide reliable forecasts of the optical turbulence above Cerro Paranal and Cerro Armazones. Such a study lead to the publication of a few papers \cite{masciadri2013, lascaux2013, lascaux2015, masciadri2017}.

\section{MODEL CONFIGURATION}

\begin{figure*}
\begin{center}
\includegraphics[angle=-90,width=0.6\textwidth]{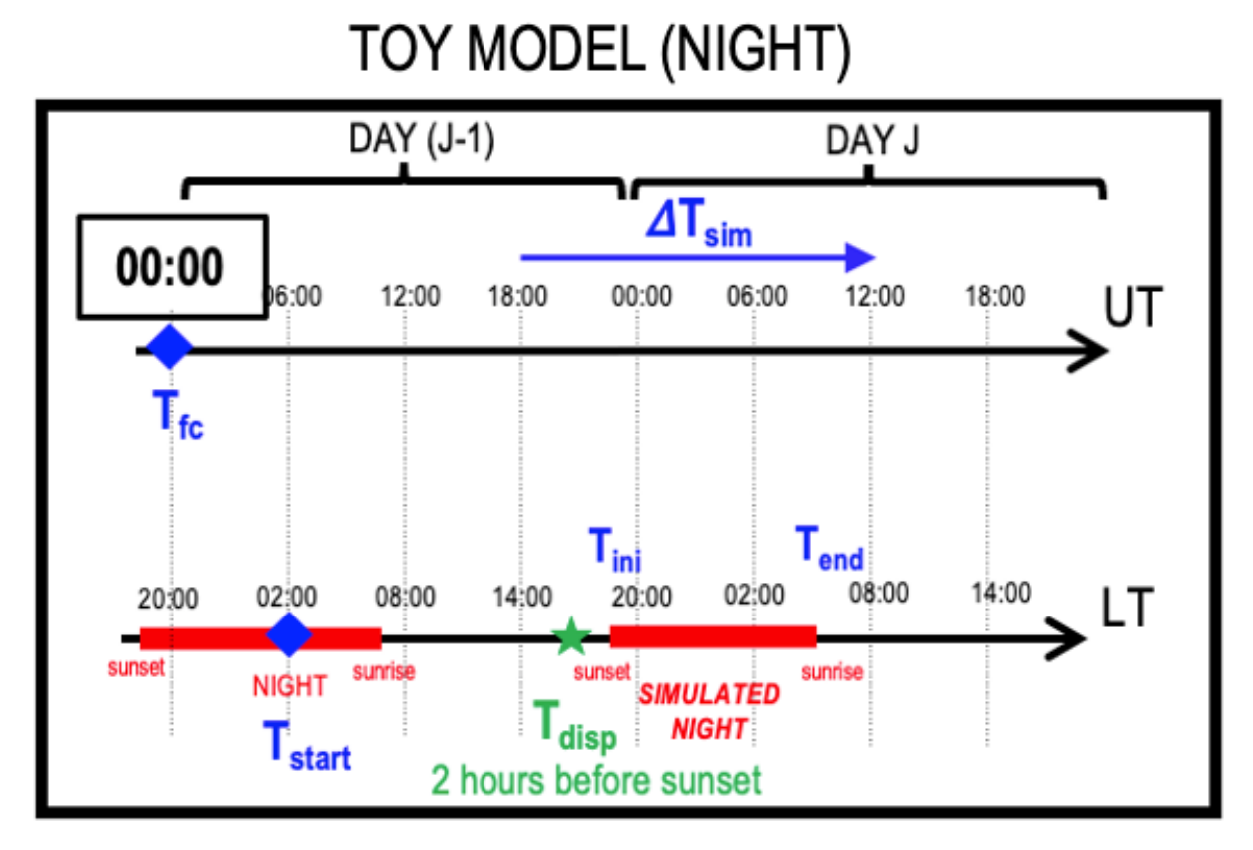}\\
\end{center}
\caption{\label{fig:fig1}  Toy model representing the configuration of the automatic forecast system related to the forecast at long time scale for 1 day. As described more extensively in \cite{masciadri2023} the forecast refers to the interval [23h - 33h]. The mesoscale model Astro-Meso-nh is initialized with data coming from the ECMWF and calculated at 00:00 UT. }
\end{figure*}

\begin{figure*}
\begin{center}
\includegraphics[angle=-90,width=0.6\textwidth]{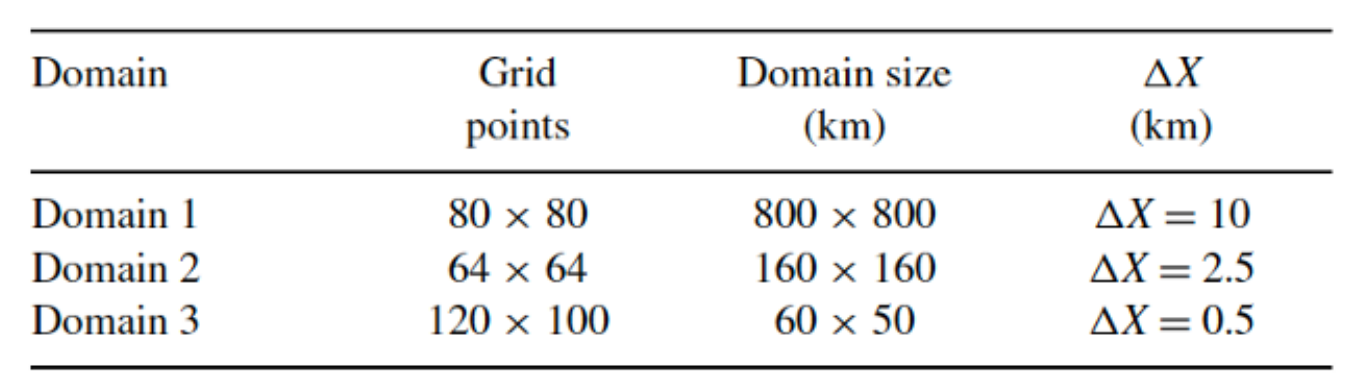}\\
\end{center}
\caption{\label{fig:fig2}  Meso-NH model grid-nesting configuration. The second column shows the number of horizontal grid points, the third column the domain extension, and the fourth column the horizontal resolution $\Delta$X .}
\end{figure*}

We refer readers to Masciadri et al. 2023\cite{masciadri2023} for a complete and extended description of the model configuration for FATE. We report here just a synthetic summary. In Fig.\ref{fig:fig1} is shown the toy model of the forecast scheme related to the forecast at a time scale of 1 day (1d) for the night time. We are considering the forecasting of the night J. The mesoscale model Astro-Meso-Nh\footnote{The Astro-Meso-Nh model is a package developed for the optical turbulence that runs jointly to the atmospheric mesoscale Meso-Nh model\cite{lafore1998,lac2018}.} is initialiazed and forced with forecasts provided by the European Centre for Medium Range Weather Forecasts (ECMWF) HRES model and calculated at 00:00 UT of the day (J-1). Such a model has 137 vertical levels and a horizontal resolution of roughly 9~km. We start the simulation with the mesoscale model Astro-Meso-Nh six hours later (at 06:00 UT) when the initialisation data are accessible and we are supposed to deliver the forecasts to ESO two hours before the sunset. To obtain the forecast at two days (2d) and three days (3d) initialization data are calculated at 00:00 UT - 24h and 00:00 UT - 48h. We highlight that the interval of time between 06:00 UT and the time of the delivery (i.e. two hours before the sunset) does not correspond to the computation time of a forecast. Actually different forecasts are executed in this interval of time. More precise and detailed infos in this respect are in Masciadri et al. 2023\cite{masciadri2023}. 
The toy model related to the day time is the same but it is shifted of 12 hours. That means that, for the forecast of the day J, initialization data from ECMWF are calculated at 12:00 UT and the simulations starts at 18:00 UT of the day (J-1). In that case we are supposed to deliver forecast two hours before the sun-rise. 
We use a grid-nesting technique\cite{stein2000} centred above Cerro Paranal (i.e. the point of interest). We have three imbricated domains and the highest horizontal resolution of the innermost domain is equal to 500~m (Fig.\ref{fig:fig2}).

\noindent
It is important to highlight that the configuration of the automatic forecast system selected for FATE is the result of a trade-off between:\\
- the number of parameters that have to be forecasted\\
- technical specifications requested by ESO\\
- hardware cost  \\
- the operational configuration that depends on the ESO requests therefore they depends on the Statement of Work (SoW).

Two different products are produced by FATE:\\

1. Forecasts at long forecast time scale (FTS): at 1 day (1d), 2 days (2d) and 3 days (3d). These three forecasts correspond to the following time scales (23h - 33h), (47h - 57h) and (71h - 81h) respectively. For completeness, the FTS for the 1d case is calculated as (23h - 33h) = T$_{ini}$ - T$_{fc}$ (see Masciadri et al. 2023\cite{masciadri2023} for details). For the forecasts at long time scale we use a pure hydrodynamical approach that is we use only the atmospherical Astro-Meso-Nh model. \\ 

2. Forecasts at short forecast time scale (FTS) more precisely at 1h or 2h that are also the most relevant intervals for the Service Mode of the VLT (but also of most of Observatories). We refer to Masciadri et al. 2023 - Section 3 for the calculation of the FTS in the case of shirt FTS. We use here a hybrid approach based on the autoregression technics that has been proposed in Masciadri et al. 2020\cite{masciadri2020} and that we call "AR technics". Such an approach implies the use outputs of the forecast obtained with the Astro-Meso-Nh mesoscale atmospherical model plus real-time observations related to a finite number N of nights in the past. It has been proved by Masciadri et al. 2020\cite{masciadri2020} that the optimized configuration is obtained with N=5. \\ \\

\section{TECHNICAL SPECIFICATIONS}
\label{tech_spec}

\begin{figure*}
\begin{center}
\includegraphics[angle=-90,width=0.8\textwidth]{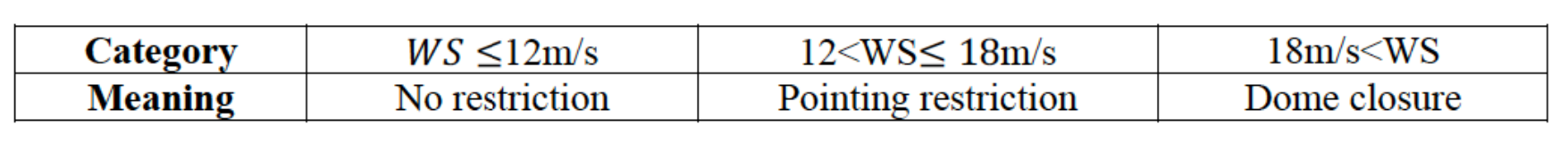}\\
\end{center}
\caption{\label{fig:fig3}  Categories, extracted from the SoW, identifying the values of the WS to be discriminated.}
\end{figure*}

\begin{figure*}
\begin{center}
\includegraphics[width=0.8\textwidth]{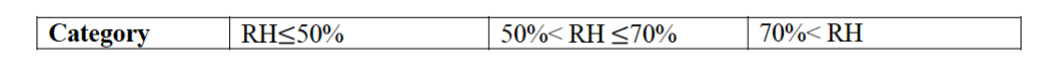}\\
\end{center}
\caption{\label{fig:fig4} Same as Fig.\ref{fig:fig3} but for the RH. }
\end{figure*}

\begin{figure*}
\begin{center}
\includegraphics[angle=-90,width=0.8\textwidth]{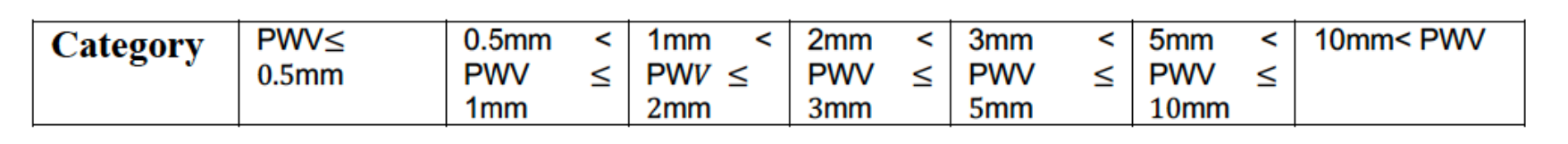}\\
\end{center}
\caption{\label{fig:fig5}  Same as Fig.\ref{fig:fig3} but for the PWV.}
\end{figure*}

\begin{figure*}
\begin{center}
\includegraphics[angle=-90,width=0.7\textwidth]{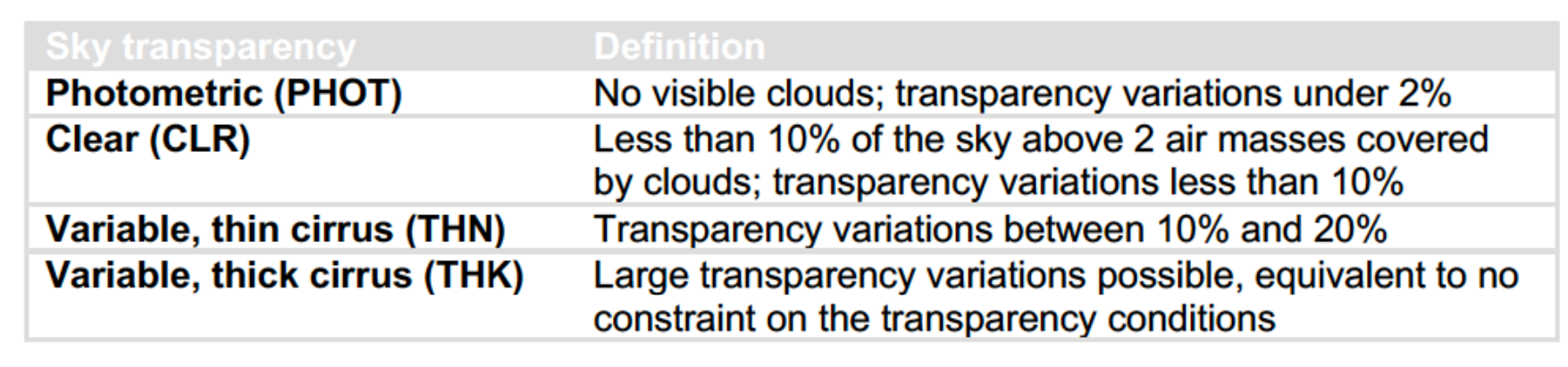}\\
\end{center}
\caption{\label{fig:fig6}  Same as Fig.\ref{fig:fig3} but for the sky transparency. }
\end{figure*}

We summarise here the technical specifications defined by ESO in the SoW. Fig.\ref{fig:fig3}-\ref{fig:fig5} show the categories within which the atmospherical parameters values for WS, RH and PWV should be distinguished. In the case of WS, when it flows with values between 12~ms$^{-1}$ and 18~ms$^{-1}$ there are pointing restrictions, when the the wind velocity is larger than 18 ~ms$^{-1}$ the dome is closed. In the case of RH, the 50\% value is related to a few humidity-sensitive instruments, 70\% for all the other. Also ESO defined a subjective classification of the sky transparency (Fig.\ref{fig:fig6}). We have four categories: the photometric sky, the clear sky, the variable thin cirrus and the variable thick cirrus. Such a classification has been proven to be strictly correlated to the cloud cover\footnote{We have no reference in this respect. We just note that this has been part of the activity/studies performed at ESO in the past and it integral part of the SoW.} therefore we provide a forecast of the sky transparency passing by the forecast of the cloudiness fraction. In it planned by ESO in the future to introduce a classification of the sky transparency based on the fluctuation analysis of the infrared sky temperature measured by the radiometer LHATPRO\cite{kerber2016}.

\begin{figure*}
\begin{center}
\includegraphics[angle=-90,width=0.9\textwidth]{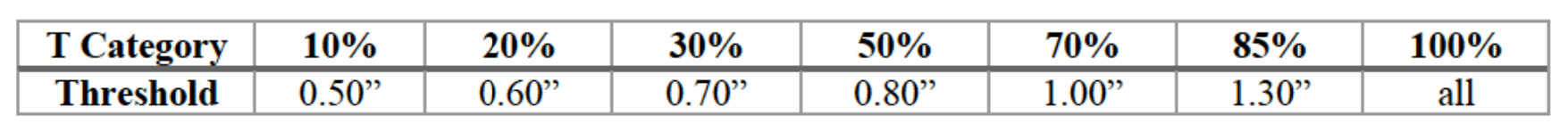}\\
\end{center}
\caption{\label{fig:fig7}  Same as Fig.\ref{fig:fig3} but for the seeing. }
\end{figure*}

\begin{figure*}
\begin{center}
\includegraphics[angle=-90,width=0.9\textwidth]{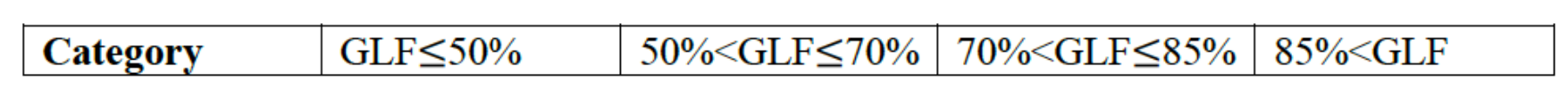}\\
\end{center}
\caption{\label{fig:fig8}  Same as Fig.\ref{fig:fig3} but for the GLF. }
\end{figure*}

\begin{figure*}
\begin{center}
\includegraphics[angle=-90,width=0.9\textwidth]{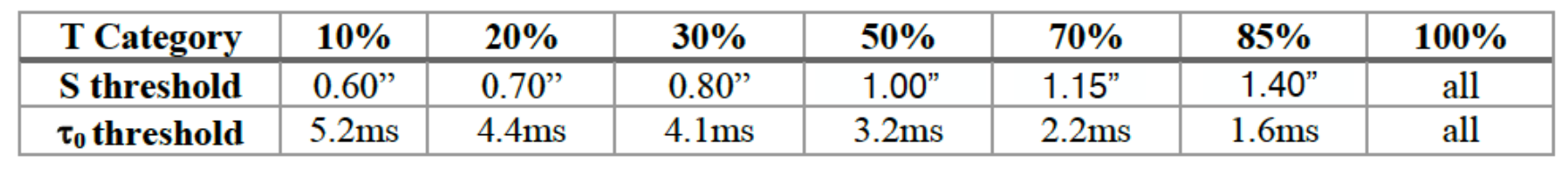}\\
\end{center}
\caption{\label{fig:fig9}  Same as Fig.\ref{fig:fig3} but for the seeing and the $\tau_{0}$. }
\end{figure*}

\begin{figure*}
\begin{center}
\includegraphics[angle=-90,width=0.7\textwidth]{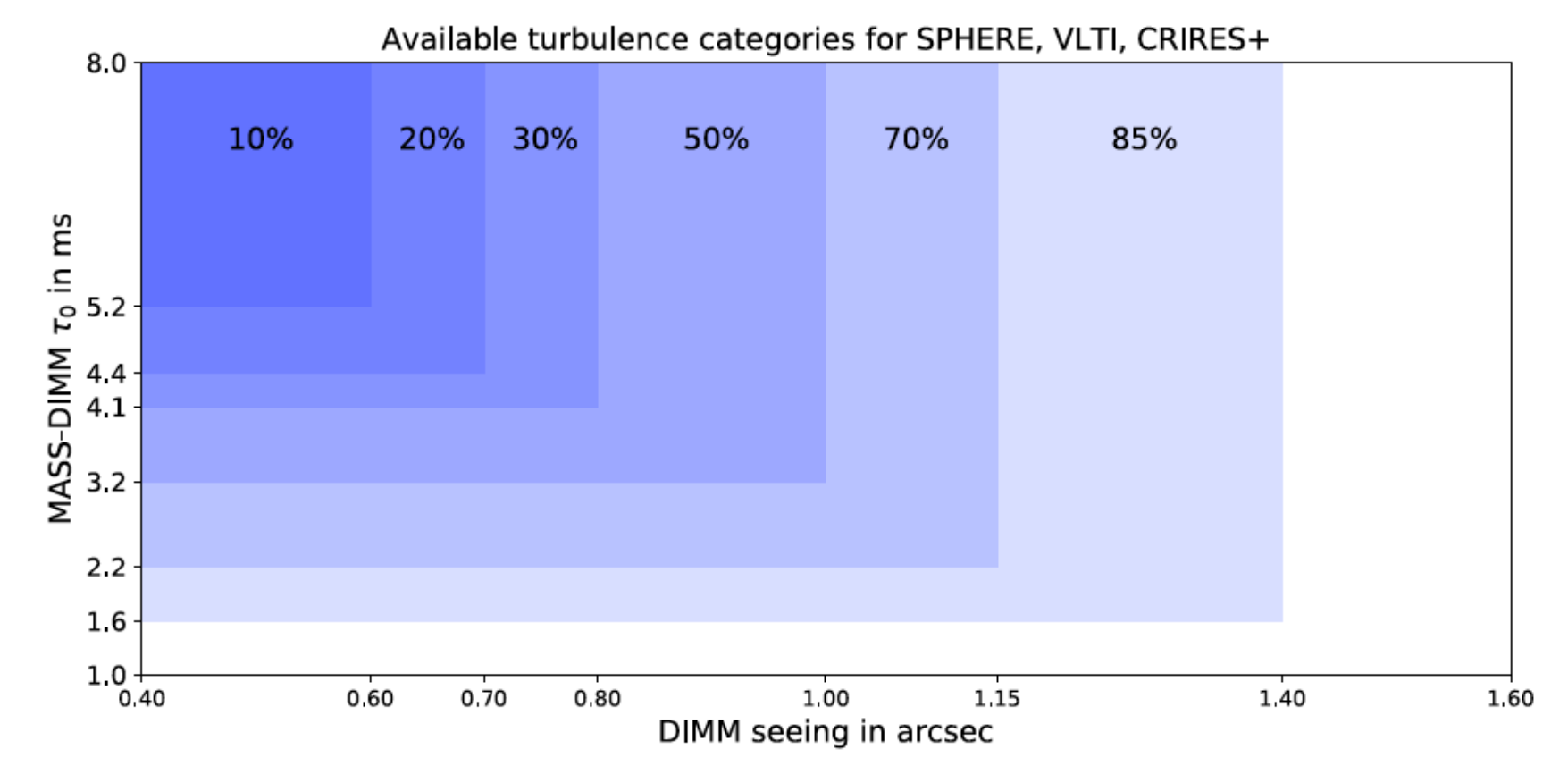}\\
\end{center}
\caption{\label{fig:fig10} Table reflecting the joint categorisation for seeing and $\tau_{0}$ shown in Fig.\ref{fig:fig9}. }
\end{figure*}

In Fig.\ref{fig:fig7} and Fig.\ref{fig:fig8} we have the categorisation for the seeing and the GLF. In Fig.\ref{fig:fig9}-Fig.\ref{fig:fig10} is shown the joint categorization for seeing and $\tau_{0}$ that are thresholds of seeing and $\tau_{0}$ for which seeing and $\tau_{0}$ are minor or equal to 10\%, 20\% ect. of the cumulative distribution calculated on a climatological scale. We highlight the fact that requirements of ESO are very, very strict. Indeed, looking at Fig.\ref{fig:fig7}  we can see that the requested accuracy can arrive for the seeing up to 0.1" in the range [0.5" - 0.8"] and looking at Fig.\ref{fig:fig9} the accuracy of $\tau_{0}$ can be smaller than 1 ms. This is because only one instrument is taken as a reference (the MASS-DIMM in this case).

\begin{figure*}
\begin{center}
\includegraphics[angle=-90,width=0.6\textwidth]{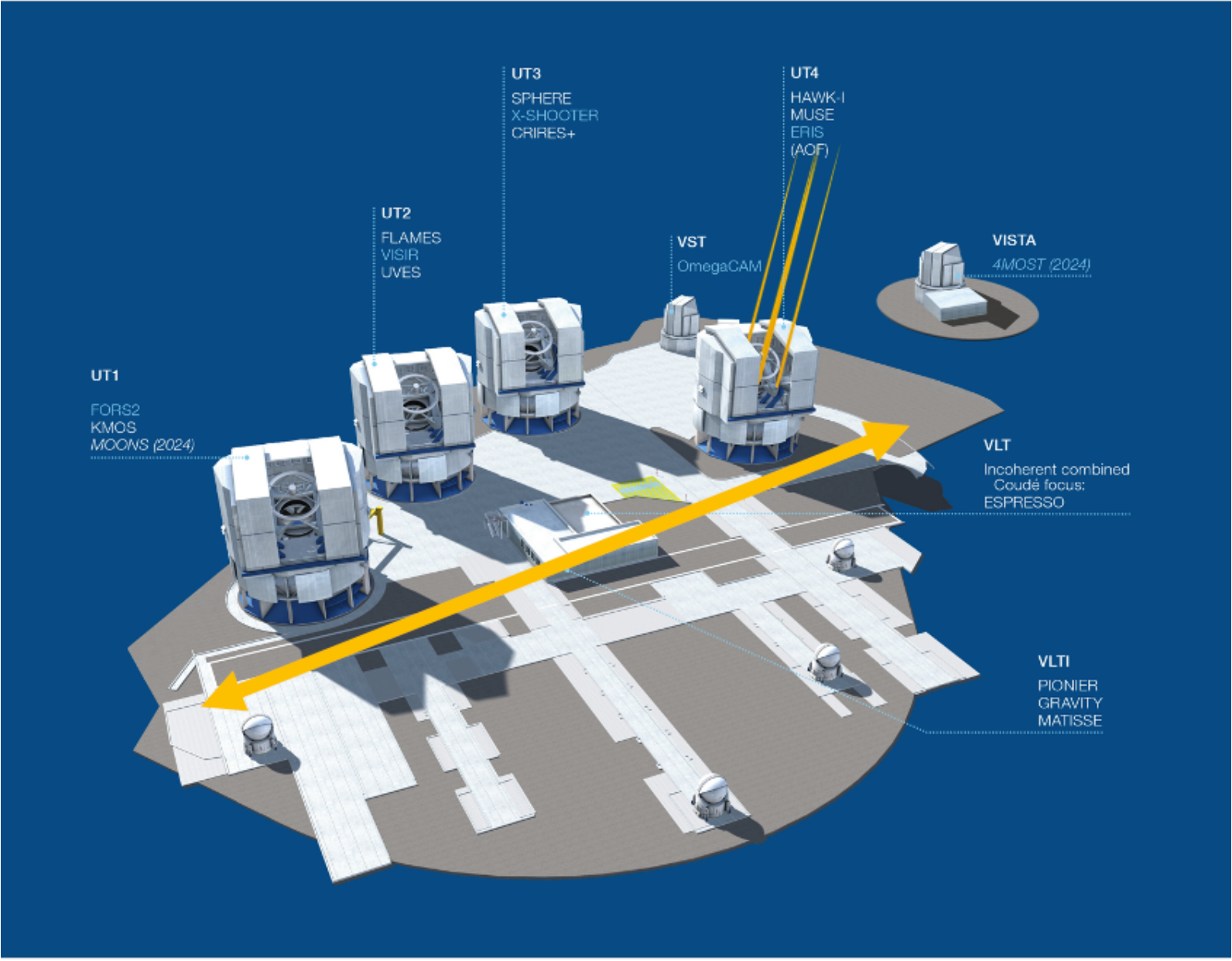}\\
\end{center}
\caption{\label{fig:fig11} Cerro Paranal plateau with the four UTs. }
\end{figure*}

On the other side, there is no evidence, at present, that these accuracies can be reached by observations. Indeed, for example, Masciadri et al. 2023\cite{masciadri2023} showed that, on a sample of 157 nights, the intrinsic dispersion SD$_{obs}$ (i.e. RMSE at which we rested the bias) between observations from a Stereo-SCIDAR and a MASS-DIMM is of the order of 0.24" for the seeing and 1.22 ms for the $\tau_{0}$. Both values are larger than the accuracy requested by ESO. Other authors showed even larger uncertainties. For example Griffiths et al. 2024\cite{griffiths2024} measured an uncertainty of 0.28" for the seeing and 2.21 ms for $\tau_{0}$ between two other instruments (RINGSS and 24HSHIMM) located basically in the same place and at the same height above the ground. Considering that FATE forecasts are supposed to be used for the Service Mode of the four UTs (and annexed instruments) and the VLTI extended on a surface of one or two hundreds meters as shown in Fig.\ref{fig:fig11} this element will have to be taken into account at a certain point. 

\begin{figure*}
\begin{center}
\includegraphics[angle=-90,width=0.9\textwidth]{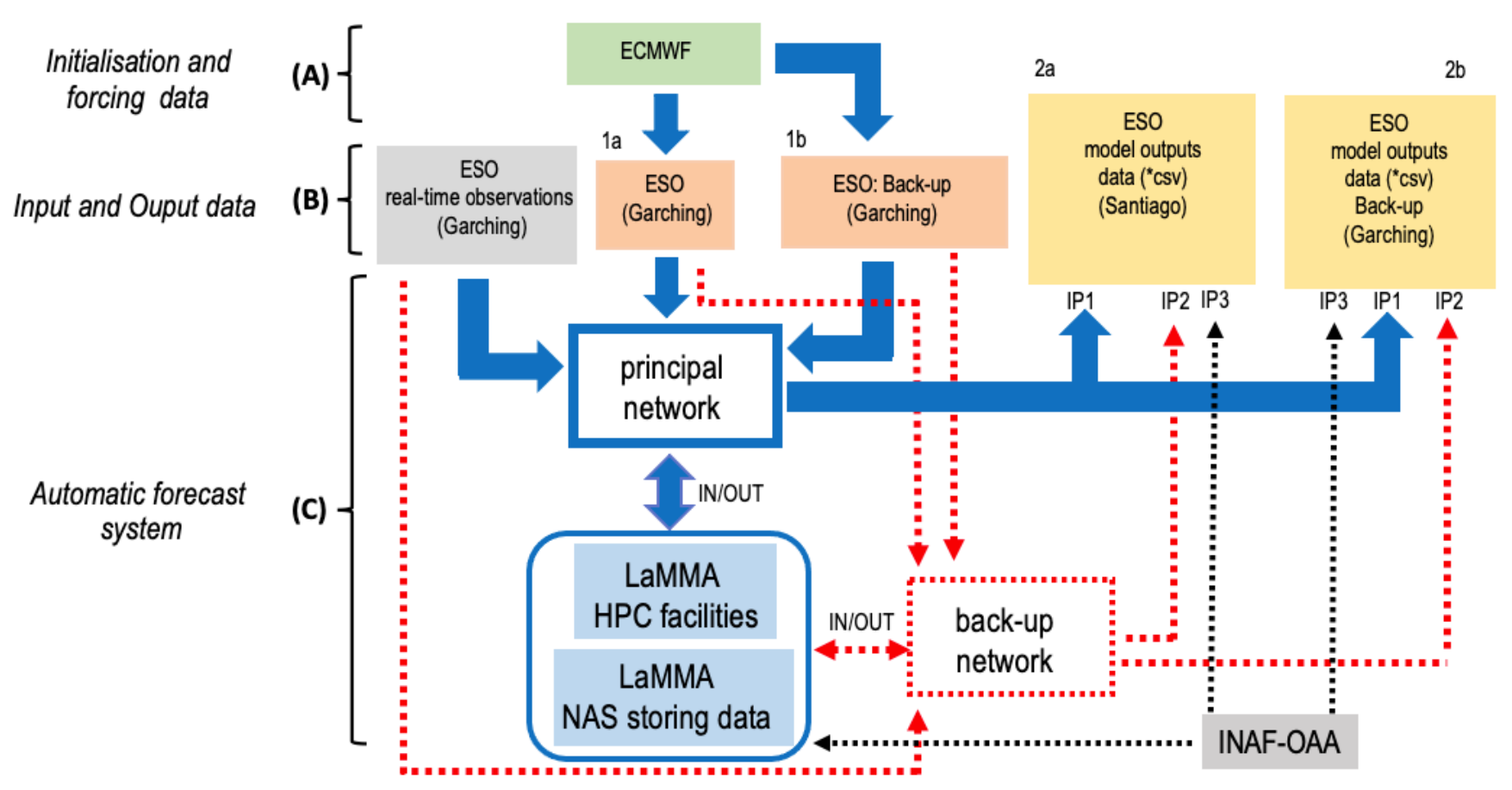}\\
\end{center}
\caption{\label{fig:fig12} Automatic forecast system of FATE.}
\end{figure*}

\begin{figure*}
\begin{center}
\includegraphics[angle=-90,width=\textwidth]{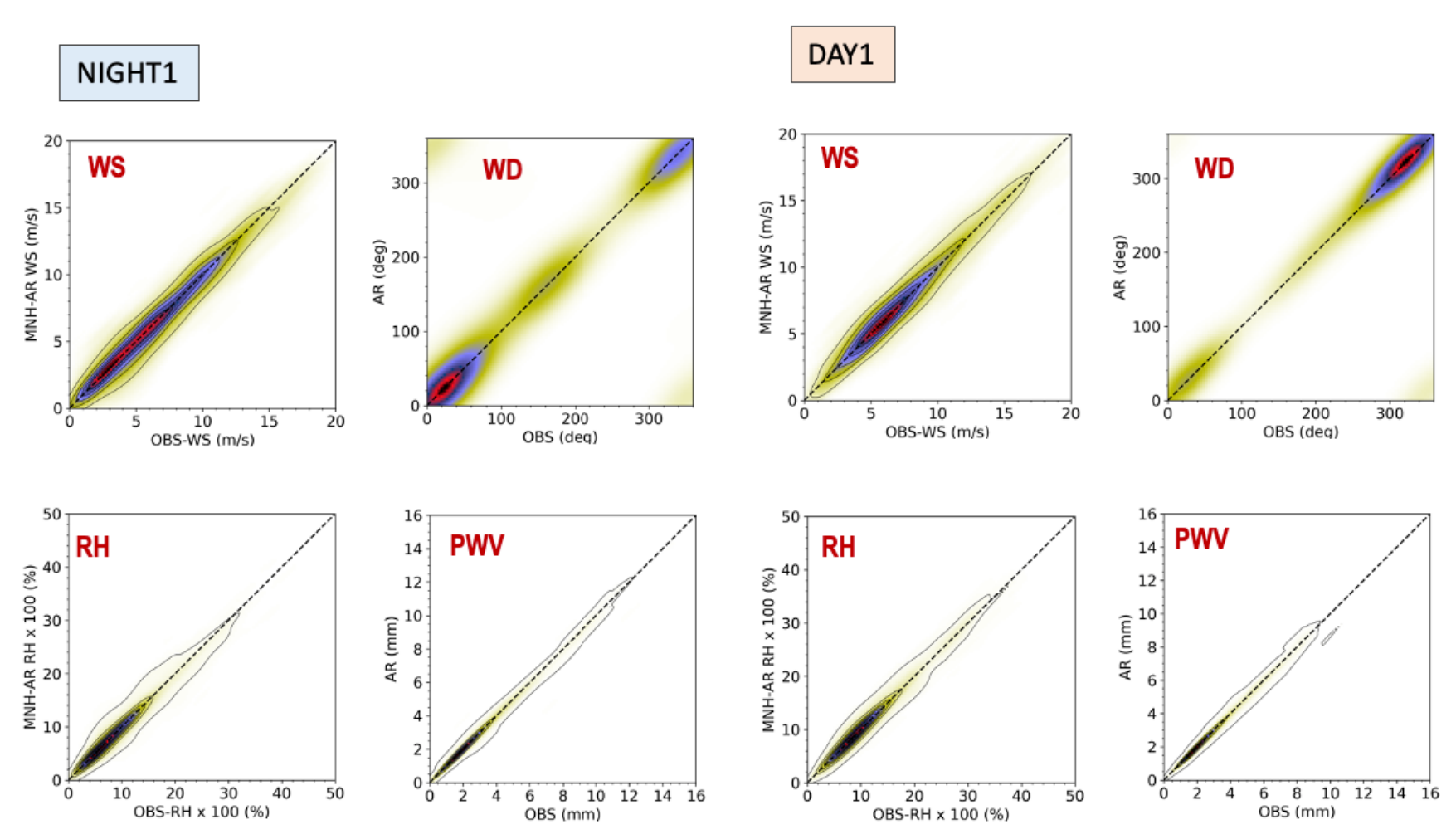}\\
\end{center}
\caption{\label{fig:fig13}  Density function maps associated to the scattering plots obtained with forecasts and observations of WS, WD, RH and PWV during the night (left) and day time (right). Sample of 4 months related to the commissioning data-set.}
\end{figure*}

\begin{figure*}
\begin{center}
\includegraphics[angle=-90,width=\textwidth]{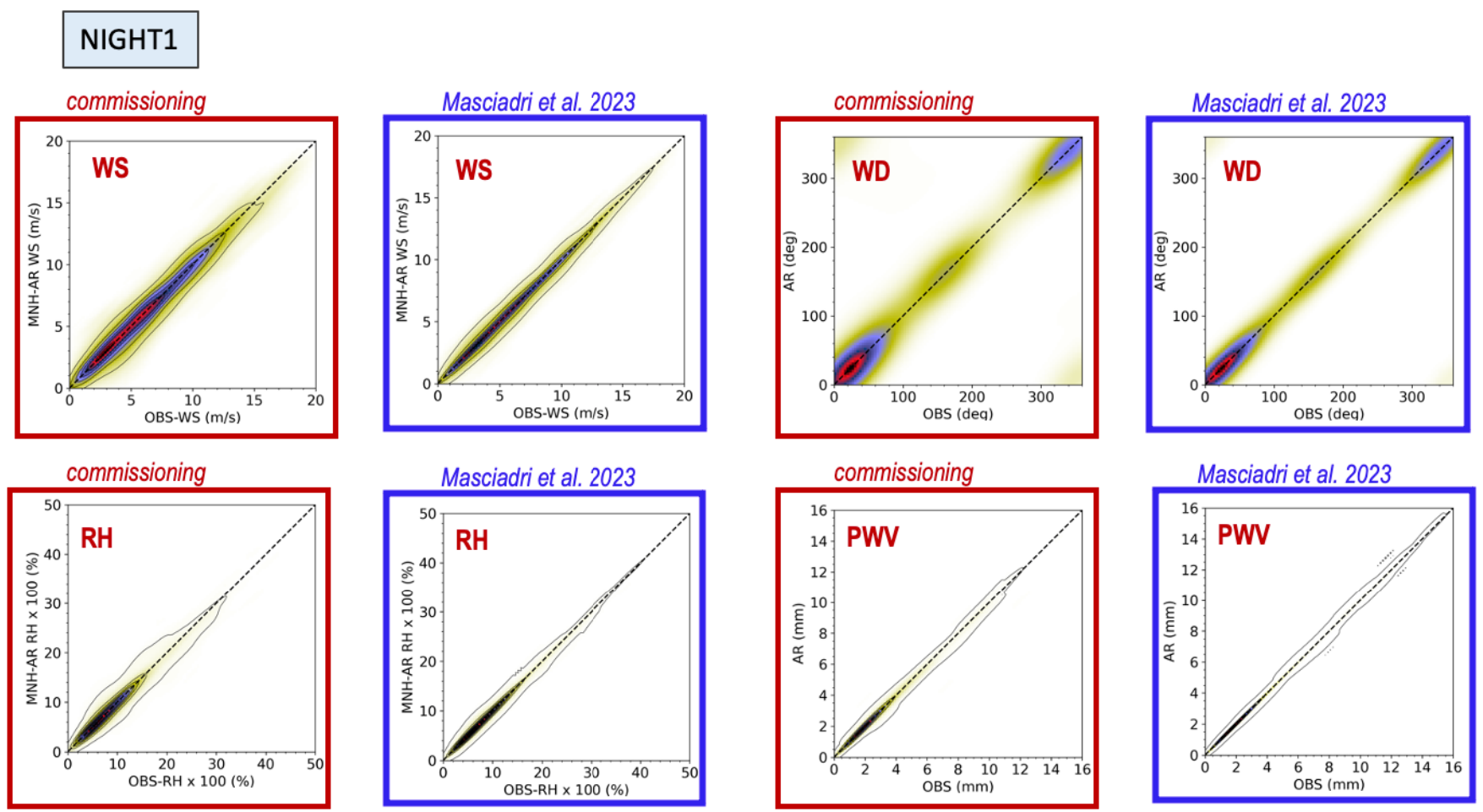}\\
\end{center}
\caption{\label{fig:fig14} Density function maps of WS, WD, RH and PWV related to the night time. The red square report results of the forecasts obtained during the commissioning. In blue squares results as treated following Masciadri et al. 2023\cite{masciadri2023} study. Sample of 4 months related to the commissioning data-set.}
\end{figure*}

\begin{figure*}
\begin{center}
\includegraphics[angle=-90,width=\textwidth]{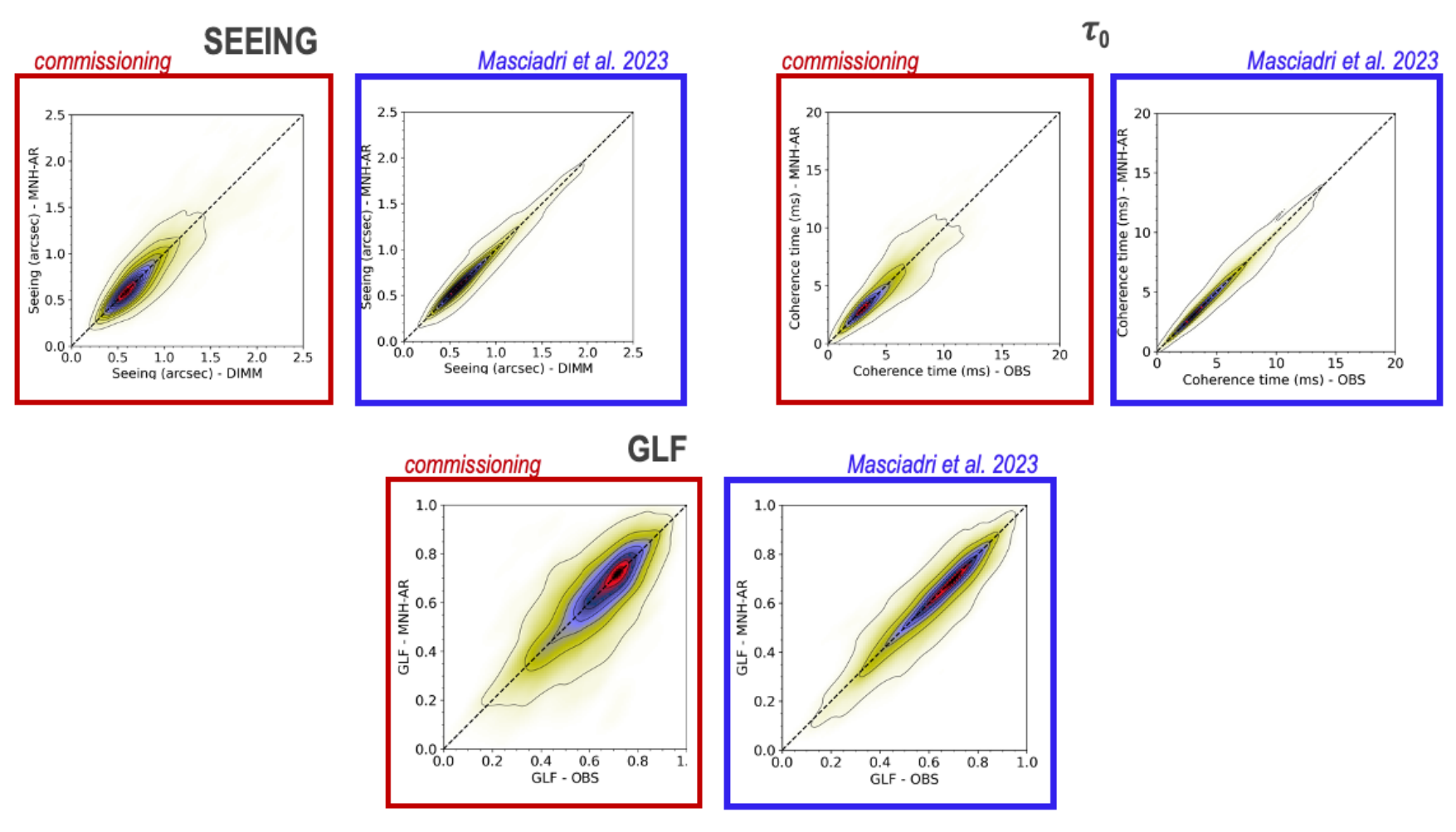}\\
\end{center}
\caption{\label{fig:fig15} As in Fig.\ref{fig:fig14} but for seeing, $\tau_{0}$ and GLF. In this picture we are considering all the values of the seeing without any filtering. }
\end{figure*}

\section{PARAMETERS/DELIVERABLES}

Following the Statement of Work (SoW) of the CfT, the parameters for which ESO required forecasts and the characteristics of the deliverables are the following:\\ \\
\noindent
NIGHT TIME\\
- Forecasts at long time scale are at 1d, 2d and 3d.\\
- We are considering forecasts of eight parameters: the wind speed (WS) at 30 m above the ground level (a.g.l.), the wind direction (WD) at 30~m a.g.l., the relative humidity (RH) at 30~m a.g.l., the precipitable water vapour (PWV), the sky transparency plus three astro-climatic parameters i.e the seeing ($\varepsilon$), the wavefront coherence time ($\tau_{0}$) and the ground layer fraction (GLF).\\
- Forecast at short time scale. ESO requires upgrades of the forecast at a frequency of 1h.\\
- The forecast temporal sampling is 10 minutes. We highlight that General Circulation Models (GCMs) can not provide such a kind of products.\\ \\
\noindent
DAY TIME\\
The specifications for the day time are the same of the night time but we are not supposed to provide forecasts of the three astroclimatic parameters. \\ 

Among the deliverables we mention also the probability that individual parameters fall in the categories defined by ESO in the SoW. In case the categories are joint categorisation (see for example Fig.\ref{fig:fig9}-Fig.\ref{fig:fig10}), we provide a joint probability.

\section{AUTOMATIC FORECAST SYSTEM}

The automatic forecast system connects institutes located in different places in the world. Initialisation and forcing data are delivered by ECMWF to ESO. These data are retrieved from ESO and computations of the forecasts with the mesoscale model are performed by FATE in Florence, more precisely in the LaMMA buildings where are located the HPC facilities and the NAS servers for the data storage. Finally, forecasts are delivered to ESO-Santiago (main delivery server) and ESO-Garching (back-up). In Fig.\ref{fig:fig12} is shown the complete flowchart/structure of the automatic forecast system. Such a system is divided in three sectors: (A) initialisation and forcing data, (B) input and output data and (C) automatic forecast system. The important thing to retain is that, for each step, we envisaged back-up servers and also back-up networks. At least in the part of the flowchart that depends on FATE (see Section \ref{less_lear} considerations for more infos).This to maximise the temporal coverage and minimise the possibility of failures. ESO indeed established a system of penalties in case forecasts are not delivered in duly time. The automatic system includes a continuous monitoring of the whole flowchart of information to identify the origin of possible failure. Finally, as shown in Fig.\ref{fig:fig12}, INAF can access both the LaMMA structure and ESO delivering servers.

\section{TIMELINE}

Here the complete timeline:\\
- On 2019/07/19: ESO publishes a Request for Information (RfI)\\
- On 2020/01/07: ESO publishes the Call for Tender (CfT) that implies 2 selection phases\\
- On 2020/02/26: deadline of the CfT. We have afterwards 7 months of stop due to COVID. The VLT was closed in that time.\\
- 2020/09/24: project FATE is pre-selected for phase 2 that implies the participation to a "validation test"\\
- [2021/01/11 - 2021/03/11]: duration of the validation test aiming to prove the ability of the teams to carry on such a project in operational phase\\
- 2021/11/08: ESO designates INAF winner of the CfT\\
- 2022/10/26: FATE contract is signed\\
- 2022/11/01: FATE development phase starts\\
- 2023/07/31: FATE developments phase ends\\
- [2023/09/01 - 2023/12/31]: duration of the commissioning \\
- 2024/04/31: signature of an amendment requested by ESO \\
- 2024/06/01: operational phase starts

\section{PRELIMINARY RESULTS}

We report here just a few results related to the forecasts at short time scale (AR) as these are the most interesting and critical one for the Service Mode. In Fig.\ref{fig:fig13} are shown the density function maps related to the associated scattering plots for the 4 months of data of the commissioning. We can notice that there is a very good correlation between forecasts and observations in both cases (night and day). However results and under our expectation as shown in Fig.\ref{fig:fig14} in the night time case. In the latter figure we report the comparison of results obtained with the commissioning data-set and results as expected from the analysis/study performed in Masciadri et al. 2023\cite{masciadri2023}. Similarly, in Fig.\ref{fig:fig15} are shown the density function maps of the seeing, $\tau_{0}$ and GLF related to the commissioning data-set (see density function maps inside red squares). It is possible to observe that the correlation between forecasts and measurements is very good\footnote{In this picture we are considering all the values of the seeing without any filtering. } and the uncertainty between forecasts and observations is already smaller than that obtained with observations (see digression in Section \ref{tech_spec}). However performances are weaker than those expected therefore those obtained with the treatment presented in Masciadri et al. 2023\cite{masciadri2023} study applied to the same sample of nights. The positive thing is that the origin of the problem has been identified and we are working to overcome that. 

\section{LESSONS LEARNED}
\label{less_lear}

On 2024 May 17th on the web page of ESO appeared a message that recited {\it "...Due to an important system up-grade several of ESO communication systems are unavailable..."}. This problem has been experienced because ESO had planned an upgrade of the network security system. As a result however ESO was basically not reachable for one week. The consequence on FATE was that it remained completely stuck for the whole duration of time. More precisely:\\
- initialisation and forcing data from ECMWF could not be delivered to ESO servers\\
- ESO archive observations useful for AR techniques related to the short FTS were unreachable \\
- FATE could not deliver forecasts to ESO servers

Of course it is absolutely normal and understandable that ESO takes care to protect its network and that means that, even if they are rare, these events may occur during the time. Solutions to this problem can be envisaged and have been discussed even if it is not suitable to deal about that in this context for security reasons. 

In terms of lessons learned suitable to be highlighted we also mention a modified way to store measurements of the PWV in the ESO archive during the night time that we suggested to implement as it would improve the performances of the forecasts at short time scale of that parameter. This has been already discussed with the ESO counterpart and approved. At present, indeed, the frequency of measurements of the PWV done at at zenith and stored in the ESO archive during the night time is lower that that during the day time. This is due to the necessity to perform measurements at different lines of sight as it is important the calibration of astronomical observations with respect to the telluric lines that requires very precise measurements of
the PWV at different lines of sight. However it has been observed\cite{querel2014} that the PWV horizontal distribution is quite
homogeneous. Spatial variability has been documented to be of the order of rms = 3\% (implying that for a 3~mm PWV, the rms is below 0.1~mm). We therefore suggested to store in the ESO archive measurements done along whatever lines of sight but corrected by the airmass [using the elevation angle 1./cos(90-ElevationAngle)]. This should guarantee a high frequency of observations of 1-2 min that is necessary for us to provide performances as shown in \cite{turchi2019,turchi2020}.

\section{CONCLUSIONS}

We can say that the FATE development phase has been concluded within the scheduled program. The commissioning lasted four months [2023/09/01 - 2023/12/31] showing no major problems and it led:\\
- to prove the strengthness of the automatic forecast system that revealed to be very robust (100\% reliability i.e. no failures);\\
- to produce an automatic report to be delivered to ESO with monthly frequency containing the state of art of the success rate of the forecast system and the percentage of ESO real-time observations availability;\\
- performances of forecasts at long time scales revealed to be those expected;\\
- performances of forecasts at short time scales showed an uncertainty with respect to observations that is better than the experienced dispersion of observations from different instruments. However performances are lower than what expected. More precisely, the RMSE is larger than that obtained following the treatment presented in Masciadri et al. 2023\cite{masciadri2023}. The problem has been identified and we are working to overcome that; \\
- we identified a few lessons learned at conclusion of the commissioning phase;\\
- Operational Phase started on 2024 June 1st.\\ \\
For what concerns the next steps, we can synthetically say that:\\ \\
- the next milestone/verification is planned for June 2025 \\
- we are engaged on R\&D to improve forecasts performances and to investigate machine learning approaches \\
- we planned to up-grade soon the forecasts at short time scales (AR approach) by changing the frequency of the up-grade forecasts to 10~min instead of 1 hour.

\acknowledgments 
 
The study is funded by the contract FATE N. PO102958/ESO/20/95952/FLAB. We acknowledge the Tuscany Region for the contribution to this activity. Initialisation data of the Astro-Meso-Nh model come from the HRES model of ECMWF. We acknowledge a few members of ESO staff for useful discussions: Alain Smette, Miska Le Louarn, Michele Cirasuolo, Olivier Hainaut, Joseph Anderson.

\bibliographystyle{spiebib} 

\begin{thebibliography}{1}

\bibitem{silva2002}
Silva, D.R., {\em SPIE Proceedings 4844}, 94

\bibitem{anderson2024}
Anderson, P.A., Sedaghati, E., Cikota, A., Behara, N., Otarola, A, Steffen, M.,  2024, {\em SPIE - Observatory Operations: strategies, processes, systems X}, 13098-6

\bibitem{masciadri1999}
Masciadri, E., Vernin, J., Bougeault, P., 1999, {\em A\&ASS}, {\bf 137}, 185

\bibitem{masciadri2001}
Masciadri, E. \& Jabouille, P., 2001, {\em A\&A}, {\bf 376}, 727

\bibitem{masciadri2006}
Masciadri, E. \& Egner, S., 2006, {\em PASP}, {\bf 118}, 1604

\bibitem{lascaux2011}
Lascaux, F., Masciadri, E., Hagelin, S., 2011, {\em MNRAS}, {\bf 411}, 693

\bibitem{hagelin2011}
Hagelin, S., Masciadri, E., Lascaux, F., 2011, {\em MNRAS}, {\bf 412}, 2695

\bibitem{masciadri2013}
Masciadri, E., Lascaux, F., Fini, L., 2013, {\em MNRAS}, {\bf 436}, 1968

\bibitem{masciadri2017}
Masciadri, E., Lascaux, F., Fini, L., 2017, {\em MNRAS}, {\bf 466}, 520

\bibitem{masciadri2020}
Masciadri, E., Martelloni, G., Turchi, A., 2020, {\em MNRAS}, {\bf 492}, 140

\bibitem{turchi2019}
Turchi, A., Masciadri, E., Kerber, F., Martelloni, G., 2019, {\em MNRAS}, {\bf 482}, 206

\bibitem{turchi2020}
Turchi, A., Masciadri, Pathak, P, Kasper, M., 2020, {\em MNRAS}, {\bf 497}, 4910

\bibitem{masciadri2023}
Masciadri, E.,Turchi, A., Fini, L., 2023, {\em MNRAS}, {\bf 523}, 3487

\bibitem{turchi2024}
Turchi, A., Masciadri, E., Fini, L., 2024, {\em SPIE - Software and Cyberinfrastructure for Astronomy VIII}, 13101

\bibitem{lascaux2013}
Lascaux, F., Masciadri, E., Fini, L. , 2013, {\em MNRAS}, {\bf 436}, 3147

\bibitem{lascaux2015}
Lascaux, F., Masciadri, E., Fini, L. , 2015, {\em MNRAS}, {\bf 449}, 1664

\bibitem{lafore1998}
Lafore J.-P., Stein, J., Asencio, N. et al., 1998, Annales Geophysicae, 16, 90

\bibitem{lac2018}
Lac, C. et al., 2018, Annales Geosci. Model Dev., 11, 1929

\bibitem{stein2000}
Stein J., Richard E., Lafore J.-P. et al., 2000, {\em Meteorol. Atmos. Phys.}, 72, 203

\bibitem{griffiths2024}
Griffiths, R. et al., 2024, {\em MNRAS}, {\bf 529}, 320

\bibitem{kerber2016}
Kerber, F., et al., {\em SPIE Proceedings}, {\bf 9910}, id. 99101S

\bibitem{querel2014}
Querel, R.R. \& Kerber, F., 2014, {\em SPIE Proceedings}, {\bf 9147}, 914792




\end{thebibliography}

\end{document}